# Half-Integer Spectral Zeros for Leakage Suppression in Fast Transmon Pulses

**A. M. Tishin** [a,b*]

[a] Lomonosov Moscow State University, 119991, Leninskie gory 1, Moscow, Russia

[b] Moscow Institute of Physics and Technology, 141701, Institutskiy per. 9, Dolgoprudny, Mosc. Reg., Russia

[*] tishin@amtc.org

## Abstract

A flat-top transmon pulse with cosine ramps is shown to possess exact spectral zeros at half-integer values of the product of ramp duration and anharmonicity, $\tau|\alpha| = n + 1/2$, and these zeros are shown to predict the positions of endpoint-leakage minima in the weak-drive regime. Dimensionless adiabaticity measures are widely used to characterise driven quantum systems; when such a measure is used as a scan coordinate its independence from the remaining pulse parameters must first be established. Within any self-similar one-scale envelope family $\Omega(t) = \Omega_0 f(t/\sigma)$ the non-adiabaticity parameter is exactly degenerate with pulse area, $\eta_{\mathrm{ref}}\cdot\theta = C_{\mathrm{shape}}$, so an independent ramp timescale $\tau$ is introduced to remove the degeneracy. The Fourier amplitude of the resulting envelope at the anharmonicity vanishes identically when $\tau|\alpha| = n + 1/2$, which is destructive interference between the contributions of the rising and falling edges; the cancellation is independent of the plateau length and therefore of the total pulse duration. The link to the dynamics is perturbative: to leading order the endpoint-leakage amplitude is set by the Fourier component of the product of the envelope and the instantaneous excited-state amplitude, which removes a second, plateau-derived family of envelope zeros that the dynamics does not follow. Four-level Duffing simulations with parameters representative of IBM Heron r2 devices collapse the minima onto the sequence at six total pulse durations from 50 to 100 ns with a root mean-square deviation of 0.034 in units of $\tau|\alpha|$; for the four durations resolved on a finer grid, inclusion of the first-order population weighting reduces the positional deviation from 0.0234 to 0.0096. The competing plateau condition does not collapse the data. The residual displacement from the half-integers persists in the weak-drive limit and is set by the population-weighted pulse geometry rather than by drive amplitude. The simulations follow the weak-drive sequence for $\Omega/|\alpha| \leq 0.05$ over the range tested; additional minima appear at stronger drive. Rotation-angle measurements on two IBM Heron r2 processors quantify the amplitude-calibration precision with which the predicted operating points could be addressed; no hardware measurement of the leakage minima and no gate benchmark is claimed.

## 1. Introduction

Single-qubit gates on superconducting processors are limited by the weak anharmonicity of the transmon spectrum. Shortening a gate requires a stronger and broader-band drive whose spectrum overlaps the $|1\rangle \rightarrow |2\rangle$ transition, transferring population out of the computational subspace. Because leakage is not captured by the Pauli error channels used in standard error correction analysis, it is particularly damaging for fault-tolerant operation, and its suppression has been the subject of a sustained body of work [1,2,3,4].

The established approach is envelope shaping. Derivative removal by adiabatic gate (DRAG) adds a quadrature correction proportional to the derivative of the in-phase envelope and remains the default on commercial hardware [3]. Frequency-domain generalizations extend this substantially. Fourier ansatz spectrum tuning (FAST DRAG) and higher-derivative DRAG shape the drive spectrum so as to place a null at the leakage transition, reaching a leakage error below $3.0\times10^{-5}$ at a gate duration of 6.25 ns for an anharmonicity of −212 MHz, a twenty-fold reduction relative to a conventional cosine DRAG pulse, and an error per gate of $(1.56 \pm 0.07)\times10^{-4}$ at 7.9 ns [4]. The same work identifies non-Markovian coherent errors arising from control-pulse distortion as a significant error source below 10 ns unless corrected by predistortion.

The mechanism underlying such nulls has recently been resolved into two distinct control objectives. Endpoint leakage, the population remaining outside the computational subspace at the end of the gate, is governed by the drive spectrum evaluated at the anharmonicity; transient leakage exposure accumulated during the gate depends instead on spectral weight across a finite band and governs the additional leakage induced by dephasing. A spectral null at the leakage transition therefore suppresses the endpoint amplitude without necessarily reducing transient exposure [5]. The present work computes endpoint populations only and does not address transient exposure.

The boundary of the accessible regime has also been mapped. Full multilevel time-domain simulations spanning adiabatic, intermediate and diabatic driving identify a threshold at $\Omega_{max}/|\Delta| \approx 0.15–0.20$, where $\Omega_{max}$ is the peak Rabi amplitude and Δ is the anharmonicity notation of Ref. [6] (denoted α below), beyond which gate fidelity deteriorates irrespective of waveform engineering; this is presented as a unified quantitative map linking pulse duration, drive strength and anharmonicity, and as the reason why picosecond-scale gates remain unattainable despite ideal two-level predictions [6]. A bound of this kind is empirical rather than a quantum speed limit in the strict sense: earlier work on frequency-crowded systems shows that leakage-state anharmonicity does not by itself impose a fundamental limit on gate speed [7].

On the control-error side, the leading analytic DRAG correction is set by the transmon anharmonicity [2,3]. More recent numerical studies compare the resulting error budgets and robustness tradeoffs with optimized pulses [8]. Experimentally, drive non-linearity produces rotation-angle deviations of up to several degrees from a linear amplitude model for a 15 ns pulse; the resulting control errors reach $2\times10^{-4}$ and account for half the total gate error, while leakage remains below $6\times10^{-5}$, as measured by error amplification [9]. Gates deliberately made robust to amplitude and frequency fluctuations have been derived by numerical optimal control and characterised experimentally [10].

Against this background the present work addresses a question that is logically prior. Dimensionless adiabaticity measures such as $\eta = |d\Omega/dt|/X^2$ are widely used to characterise driven quantum systems and have been introduced in earlier work as a local measure of the breakdown of the adiabatic approximation [11]. When such a measure is used as a scan coordinate rather than as a diagnostic, its independence from the remaining pulse parameters must be established first.

Section 2 shows that it cannot, within a fixed envelope shape: an exact relation fixes the product of $\eta$ and the pulse area at a constant determined by the shape alone, so that a scan in $\eta$ is a scan in rotation angle. An envelope family carrying two independent timescales is then constructed, in which the constraint is lifted.

Sections 3 and 4 present the numerical and analytic results together with the accompanying hardware calibration measurements. Section 5 states the resulting experimental-accessibility limits and what the measurements on ibm_fez and ibm_kingston can and cannot establish.

Future work will address the extension of this analysis to two-qubit gates and the inclusion of the transfer function of the control chain, which is not modelled here and which sets the practical tolerance on the operating points identified below.

## 2. Methods

The non-adiabaticity parameter is understood here as a family $\eta_X \equiv |d\Omega/dt| / X^2$, where the reference scale X is set by the physical mechanism under consideration. Throughout, $\Omega(t)$ denotes the envelope of the resonant drive, that is the instantaneous Rabi frequency in angular units, and $\omega_0$ denotes the bare qubit transition frequency; both are defined explicitly by the Hamiltonian of Section 3. Two members are relevant to the present discussion. The first, $\eta_\Omega \equiv |d\Omega/dt|/\Omega^2$, compares the drive rate to the instantaneous Rabi splitting and therefore controls the adiabaticity of the dressed states; its denominator is the instantaneous gap on resonance, and it reduces to the Landau–Zener criterion at $\eta_\Omega \sim 1$, so that the value of order unity is a definition rather than a fitted threshold. The second, $\eta_\omega \equiv |d\Omega/dt|/\omega_0^2$, compares the drive rate to the bare transition frequency and controls the weight of counter-rotating contributions, that is, the validity of the rotating-wave approximation; it becomes of order unity only when $|d\Omega/dt|$ approaches $\omega_0^2$. These are therefore not one quantity written in two ways but two dimensionless numbers governing two distinct transitions. They are related exactly by $\eta_\omega / \eta_\Omega = (\Omega/\omega_0)^2$, and coincide only in the ultrastrong-drive limit $\Omega \to \omega_0$.

For the fixed-frequency transmons considered here the Rabi rate ranges from a few megahertz to a few hundred megahertz against a transition frequency of 4.80 GHz, so $\Omega/\omega_0$ lies between about $2\times10^{-3}$ and $4\times10^{-2}$ and $(\Omega/\omega_0)^2$ between $4\times10^{-6}$ and $1.6\times10^{-3}$. The two scales therefore differ by at least three orders of magnitude, and $\eta_\Omega$ is the operative parameter throughout this work; all values of $\eta$ quoted below refer to $\eta_\Omega$. A value of $\eta_\omega$ of order unity would require a Rabi rate comparable to the transition frequency itself, far beyond the ultrastrong-drive regime and beyond the range in which the transmon remains a well-defined qubit. The second member becomes operative in platforms where $\Omega/\omega_0$ is not small, and the family is stated here in full so that the reference scale used in any given setting is explicit. Since $\eta_X$ is a local quantity, a scalar characterisation of a finite pulse requires a stated convention; the convention used here is $\eta_{ref} \equiv max_t\ |d\Omega/dt| / \Omega^2_{peak}$, evaluated over the pulse support. In the present work this parameter enters only as the dimensionless scan coordinate required for the envelope-degeneracy argument of the next subsection.

Consider a self-similar one-scale envelope family $\Omega(t) = \Omega_0 f(t/\sigma)$ with $\max_x(f) = 1$, in which a single amplitude and a single timescale are varied while the shape $f$ is held fixed, and let $I = \int f(x)\,dx$ and $D = \max_x |f'(x)|$. The pulse area is $\theta = \Omega_0\,\sigma\,I$ and $\eta_{ref} = D/(\Omega_0\,\sigma)$, so that $\eta_{ref}\,\theta = I\,D \equiv C_{shape}$, independent of $\Omega_0$ and $\sigma$ separately. For a Gaussian envelope $C_{shape} = \sqrt{(2\pi/e)} = 1.5203$; for sech and $\cos^2$ envelopes $C_{shape} = 1.5708$; for a Lorentzian, 2.0340; for a triangular envelope, 1.0000. Within such a family $\eta_{ref}$ is therefore a reparametrisation of the rotation angle: a scan in $\eta$ is a scan in pulse area, and structure observed in such a scan cannot be attributed uniquely to non-adiabaticity, because $\eta$ and the rotation angle vary in exact functional dependence. For a Gaussian envelope the values $\eta_{ref} \simeq 0.5$ and $\eta_{ref} \simeq 1$ correspond precisely to a π pulse and a π/2 pulse. The degeneracy is lifted only by an envelope family carrying two independent timescales: a flat-top envelope with cosine ramps of duration $\tau$ at each edge and total duration $T$ has $\eta_{ref} = \pi(T - \tau)/(2\tau\theta)$, so that at fixed $T$ and fixed $\theta$ the non-adiabaticity is set by $\tau$ alone while the rotation angle is held exactly constant.

No phase-coherent logic gate is benchmarked in this work, and no randomized benchmarking or process-tomography data are reported. The hardware measurements are used only to quantify state-return behaviour and drive-amplitude calibration precision. Statements below concerning gate duration therefore refer to computed leakage and to state-return probability, not to demonstrated gate fidelity.

Figure 1 shows that within any self-similar one-scale envelope family the non-adiabaticity parameter and the pulse area are locked to one another, so that the two cannot be varied independently, and that the flat-top family with an independent ramp duration removes this restriction.

The experimental data were acquired between 31 March and 21 April 2026. The hardware characteristics of the 156-qubit IBM Heron r2 processors (ibm_kingston and ibm_fez) are summarised in Table 1.

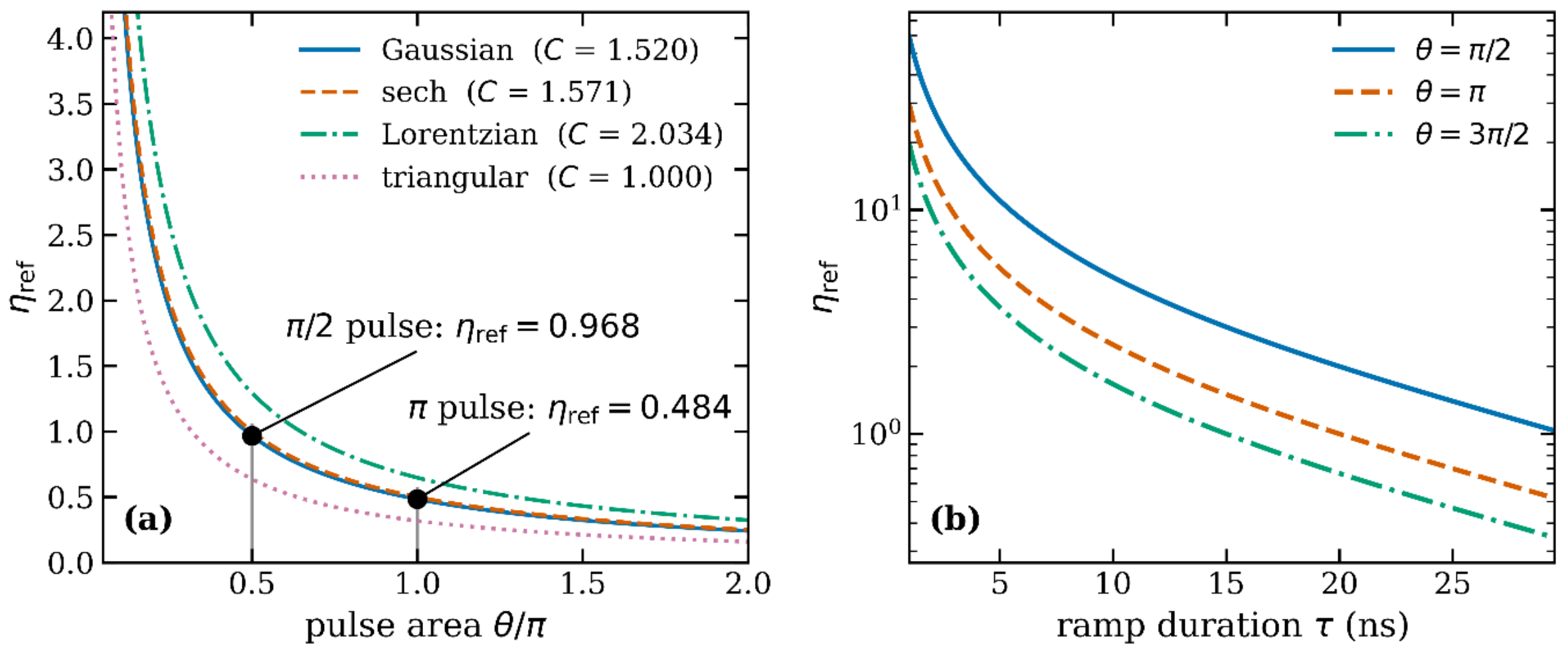

Figure 1. Degeneracy of the non-adiabaticity parameter within self-similar one-scale envelope families. (a) $\eta_{\mathrm{ref}}$ as a function of pulse area for four envelope shapes; each curve is the hyperbola $\eta_{\mathrm{ref}}\cdot\theta = C_{\mathrm{shape}}$, so that within a fixed shape the non-adiabaticity is a reparametrisation of the rotation angle. For a Gaussian envelope a π pulse corresponds to $\eta_{\mathrm{ref}} = 0.484$ and a π/2 pulse to $\eta_{\mathrm{ref}} = 0.968$. (b) The flat-top family with cosine ramps carries two independent timescales: at fixed total duration $T = 60$ ns and fixed pulse area, $\eta_{\mathrm{ref}}$ is set by the ramp duration $\tau$ alone while the rotation angle is held exactly constant.

Table 1. Device-averaged calibration parameters for the two Heron r2 processors, 17 April 2026.

| Backend | Qubit range | $T_1$ relaxation (μs) | $T_2$ dephasing (μs) | Readout assignment error |
|---|---|---|---|---|
| ibm_kingston | Q0–Q155 | 263.68 | 160.69 | $2.61\times10^{-2}$ |
| ibm_fez | Q0–Q155 | 142.37 | 104.05 | $2.74\times10^{-2}$ |

The two backends differ by nearly a factor of two in coherence time, ibm_kingston at $T_1 \approx 264$ μs and ibm_fez at $T_1 \approx 142$ μs. The drive-amplitude biases extracted in Section 3 are of comparable magnitude on both devices, indicating that these systematics are set by the control chain rather than by the relaxation rates.

Two calibration snapshots were compared to assess temporal stability. The measurements of 17 April 2026 (Table 1) gave device-averaged $T_1$ of 264 μs for ibm_kingston and 142 μs for ibm_fez; the recalibration of 20 April 2026 (Supplementary Information, Sec. S2) gave 240 μs and 48 μs respectively. On qubit 0 of ibm_kingston, $T_1$ changed from 293.5 μs to 329.8 μs between 20 and 21 April 2026, an increase of 12.3 %, with $T_2$ = 297.6 μs. These figures bound the interval over which a fixed operating point remains valid without recalibration.

## 3. Results

In the weak-drive limit the leading endpoint-leakage amplitude is controlled by the Fourier component of the drive envelope at the anharmonicity, and the calculations below examine how that component depends on the ramp duration at fixed pulse area.

The numerical model is a driven Duffing transmon truncated to four levels. In the laboratory frame, $H(t)/\hbar = \omega_0 N + (\alpha_\omega/2)N(N-1) + \Omega(t)\cos(\omega_0 t)(a + a^\dagger)$, where $N = a^\dagger a$, $\omega_0/2\pi = \nu = 4.80$ GHz, and $\alpha_\omega = 2\pi\alpha$ with $\alpha = -310$ MHz. Thus $\alpha$ denotes the anharmonicity in ordinary frequency units throughout this work, while $\alpha_\omega$ is its angular-frequency counterpart entering the Hamiltonian. This convention matches the numerical implementation: $\Omega(t)$ is the angular Rabi-frequency envelope, so the resonant rotating-frame coupling is $\Omega(t)/2$. The pulse envelope $\Omega(t)$ is the symmetric flat-top cosine-ramp envelope defined in Section 2. The time-dependent Schrödinger equation is integrated with the DOP853 explicit Runge–Kutta method using absolute and relative tolerances of $10^{-10}$. The calculations reported in the main text retain four transmon levels; convergence was checked with six levels, for which the positions of the minima are unchanged and the population above $|2\rangle$ remains below $7\times10^{-8}$. No rotating-wave truncation is imposed in the numerical integration.

Ratios written below as $\Omega/|\alpha|$ are understood in ordinary-frequency units, i.e. as $(\Omega/2\pi)/|\alpha|$; equivalently, they are the dimensionless angular-frequency ratio $\Omega/|\alpha_\omega|$.

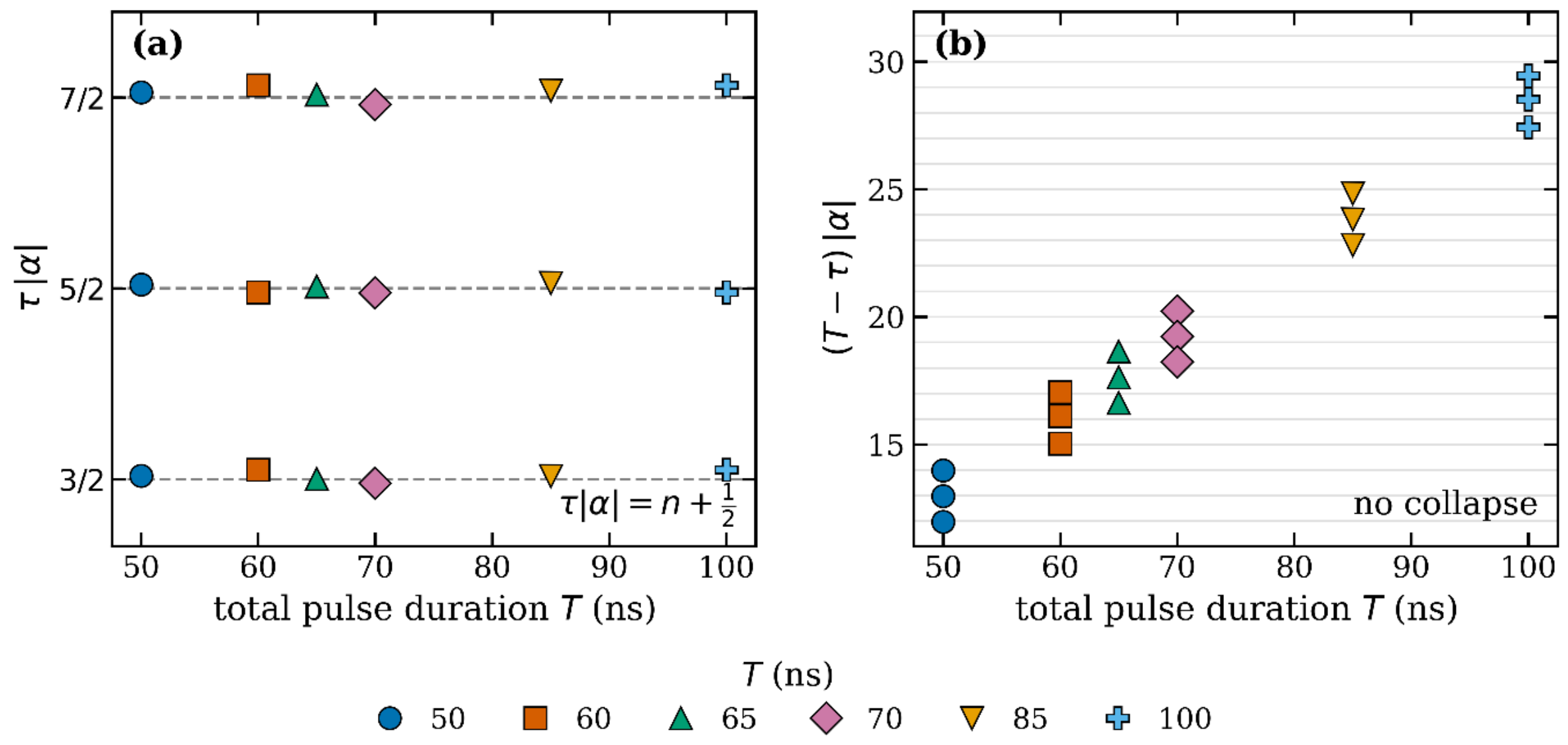

Figure 2. Position of the leakage minima for a four-level Duffing transmon ($\nu$ = 4.80 GHz, $\alpha$ = −310 MHz) at six total pulse durations between 50 and 100 ns, in the weak-drive regime $\Omega/|\alpha| \leq 0.05$. (a) Expressed as $\tau|\alpha|$, the minima collapse onto the half-integers 3/2, 5/2 and 7/2, with a root-mean-square deviation of 0.034 and a maximum deviation of 0.065. (b) The same minima expressed as $(T - \tau)|\alpha|$ show no collapse. The condition therefore constrains the ramp duration alone and is independent of the total pulse duration.

The condition $\tau|\alpha| = n + 1/2$ follows analytically from the envelope alone. Define the leakage spectral angular frequency $\varpi \equiv |\alpha_\omega| = 2\pi|\alpha|$. The flat-top envelope is symmetric about $T/2$, so its Fourier transform may be written $F(\varpi) = \exp(i\varpi T/2)\, G(\varpi)$ with $G$ real, and the zeros of $|F|$ are the zeros of $G$. Writing $a = T/2 - \tau$, $b = T/2$ and $k = \pi/\tau$, direct integration gives $G(\varpi) = 2 \sin(\varpi a)/\varpi + \cos(\varpi b) A + \sin(\varpi b) B$, with $A = \sin(\varpi\tau)[1/\varpi + \varpi/(\varpi^2 - k^2)]$ and $B = (1 - \cos \varpi\tau)/\varpi - \varpi(1 + \cos \varpi\tau)/(\varpi^2 - k^2)$. At $\varpi\tau = (2n + 1)\pi$ the first factor vanishes, $A = 0$, and the second term of $B$ vanishes with $\cos \varpi\tau = -1$, leaving $B = 2/\varpi$. The remainder is $G = (2/\varpi)[\sin(\varpi a) + \sin(\varpi a + \varpi\tau)]$, and since $\varpi\tau$ is an odd multiple of $\pi$ the second term cancels the first, so that $G \equiv 0$ for every $a$, that is for every total pulse duration T. Since $\varpi = 2\pi|\alpha|$, the condition $\varpi\tau = (2n + 1)\pi$ is exactly $\tau|\alpha| = n + 1/2$. The even case is excluded: at $\varpi\tau = 2n\pi$ one also has $A = 0$, but $B = -2\varpi/(\varpi^2 - k^2) \neq 0$ and $\sin(\varpi b) = \sin(\varpi a)$, so that $G = \sin(\varpi a)[2/\varpi + B]$ does not vanish. Integer values of $\tau|\alpha|$ are therefore not minima, in agreement with the numerical results. The two surviving terms $\sin(\varpi a)$ and $\sin(\varpi a + \varpi\tau)$ are the contributions of the rising and falling edges, separated in phase by $\varpi\tau$. The condition is thus destructive interference between the leakage amplitudes generated at the two edges, which occurs when the ramp duration contains a half-integer number of anharmonicity periods. Because the cancellation is independent of a, the plateau length and hence the total pulse duration do not enter. Numerical evaluation of $G$ confirms the derivation: at $\tau|\alpha| = 3/2$, $5/2$ and $7/2$ the analytic expression returns zero to machine precision, below $10^{-14}$, at all six total durations examined, while at integer $\tau|\alpha|$ it returns values of order $10^{-2}$.

The residual displacement of the dynamical minima from the bare half-integer sequence is not a finite-drive correction. In the weak-drive limit $c_1(t) \simeq \Theta(t)/2$, where $\Theta(t) = \int_0^t \Omega(t')\, dt'$ is the accumulated pulse area, so the relevant first-order weight is the product $\Omega(t)\Theta(t)$. After normalisation the shape of this weight is independent of the overall drive amplitude, and the displacement of the spectral zeros therefore remains finite as the drive amplitude is reduced. Numerically, at $T = 50$ ns the shift changes only from 0.0248 at $\theta = 0.25\pi$ to 0.0232 at $\theta = \pi$, an approximately elevenfold change in $\Omega/|\alpha|$, and a power-law fit over eight amplitudes returns an exponent of 0.20, excluding both linear and quadratic scaling in $\Omega$. The displacement instead depends on the pulse geometry through the total duration $T$, changes sign across the family examined, and is reproduced in sign and order of magnitude by the first-order integral. The structure follows from parity. Writing $s = t - T/2$, the envelope is even in s while the accumulated area splits as $\Theta = \theta/2 + g(s)$ with $g$ odd, so that the first-order amplitude separates into two contributions in quadrature, $A_2 \propto \exp(-i\, \bar{\omega}\, T/2)[(\theta/2)G(\tau) - iH(\tau)]$, with $H(\tau) = \int \Omega(s)g(s)\sin(\bar{\omega}\, s)ds$. The even contribution vanishes at $\tau|\alpha| = n + 1/2$; the odd one does not, so the leakage at the minimum has a nonzero floor and the true minimum is displaced. Both contributions scale as the square of the drive amplitude, which is the origin of the amplitude independence. The minimum satisfies $(\theta^2/4)GG' + HH' = 0$; at that point the two contributions are comparable in magnitude, for example $5.4\times10^{-6}$ and $4.6\times10^{-6}$ at $T = 70$ ns, so no expansion in a small parameter applies and the displacement is not available in closed form. The half-integer sequence is thus the exact spectral skeleton of the envelope, while the leading dynamical displacement is set by population weighting rather than by drive strength. The residual discrepancy between this first-order description and the full multilevel minima reflects higher-order dynamical corrections, including Stark-shift and multilevel effects.

The fine structure of the minimum near $\tau = 8$ ns, the member of the sequence with $\tau|\alpha| \simeq 5/2$, was resolved by scanning the ramp duration in steps of 0.2 ns at fixed total duration and fixed pulse area. Figure 3(a) shows the resulting profile: the leakage falls to $5.10\times10^{-9}$ at $\tau = 8.00$ ns against $2.00\times10^{-6}$ at $\tau = 6.6$ ns, a suppression by a factor of 392, and the interval within a factor of ten of the minimum is 0.40 ns wide.

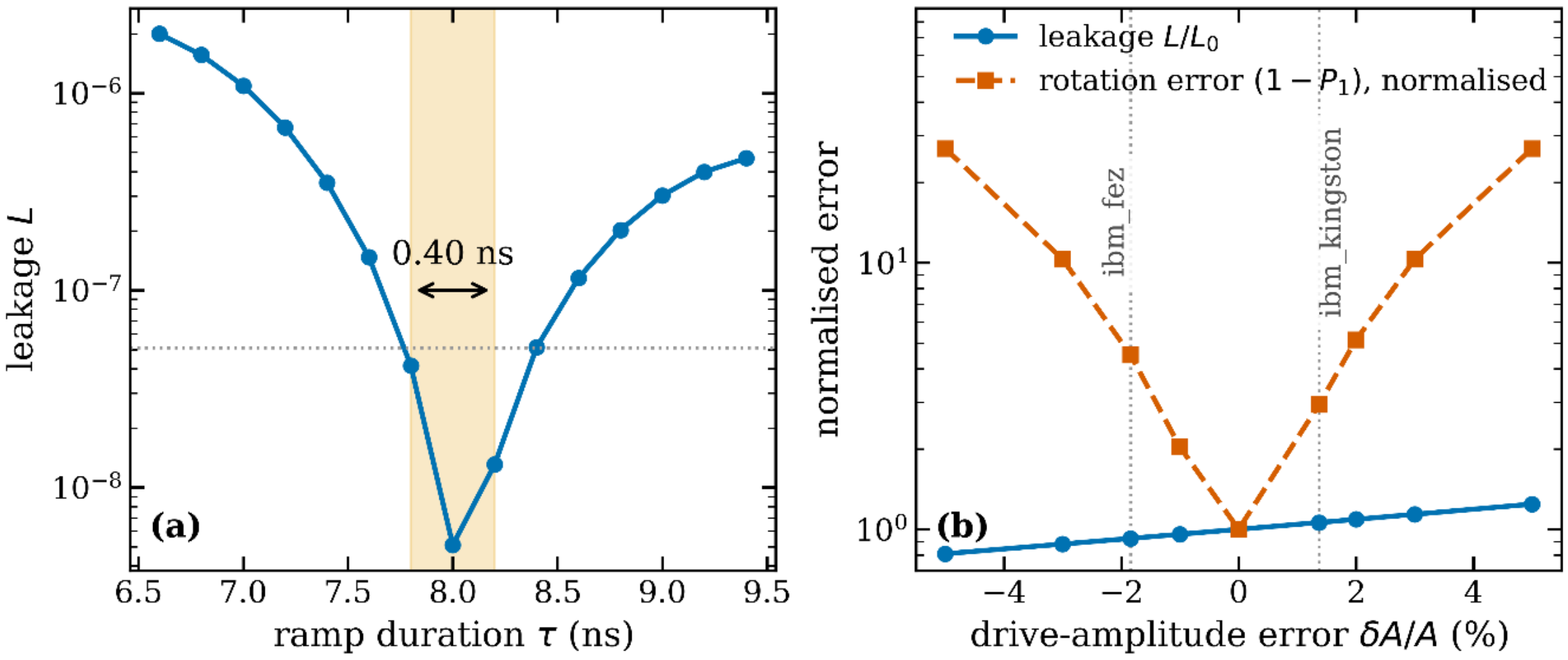

Figure 3. (a) Resolved structure of the leakage minimum with $\tau|\alpha| \simeq 5/2$, at a total duration of 60 ns and fixed pulse area $\pi$. The shaded band marks the interval within a factor of ten of the minimum, of width 0.40 ns, against a control timing granularity of 0.222 ns; the dotted line marks ten times the minimum leakage. (b) Response of leakage and of the rotation error to a systematic drive-amplitude error at the same operating point, each normalised to its value at zero error. Vertical lines mark the biases measured on the two processors. A ±5 % amplitude error changes the leakage by a factor of 1.5 and the rotation error by a factor of 27.

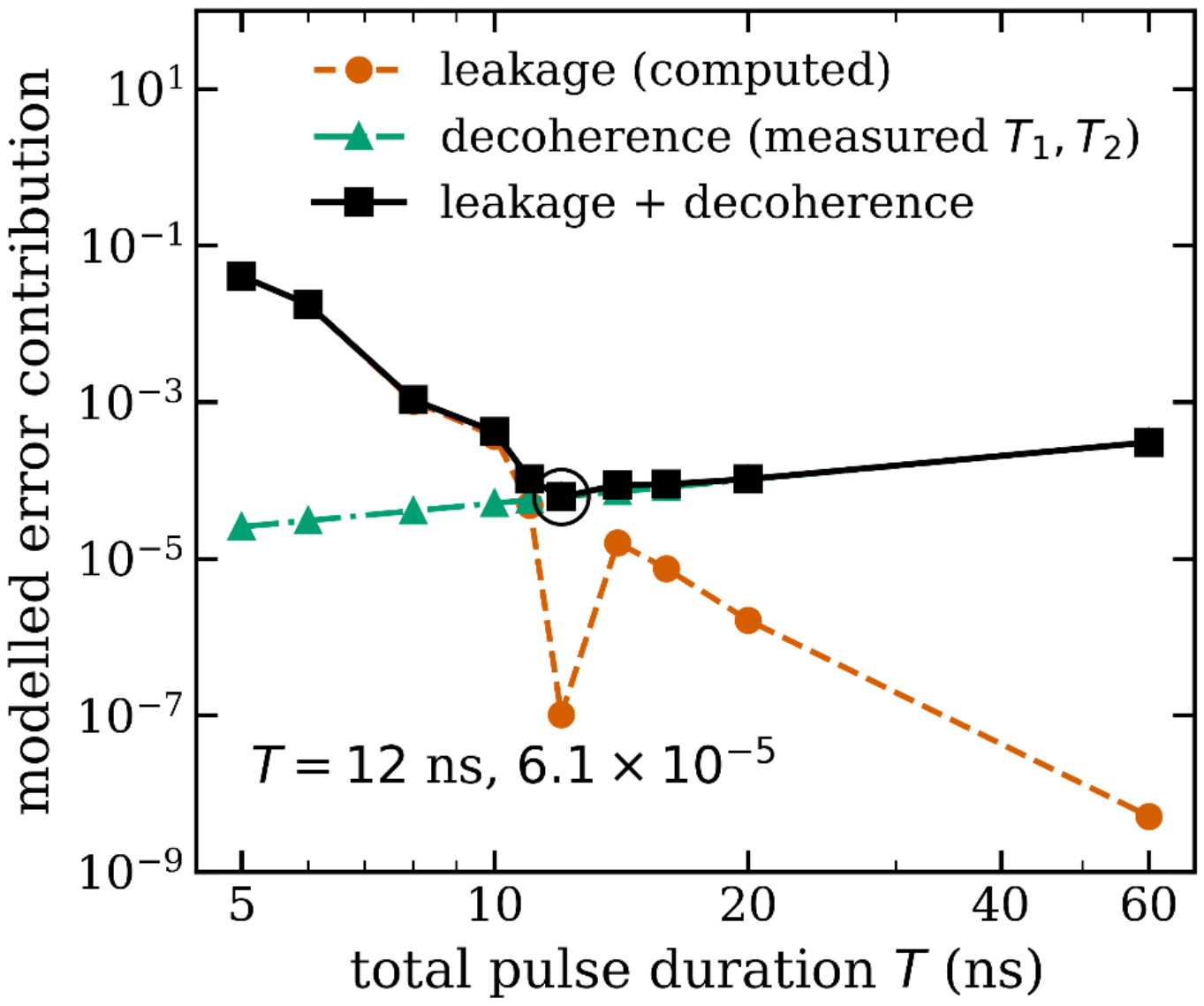

Figure 4. Modelled error budget of a $\pi$ pulse as a function of total pulse duration, comprising the computed leakage at the ramp duration that minimises it and a decoherence contribution evaluated as $T/T_1 + T/(2T_2)$ from the measured coherence times $T_1 = 293.5$ μs and $T_2 = 297.6$ μs. Control-chain distortion, coherent rotation error, drive-induced dephasing and state-preparation and measurement errors are not included, and the summed curve is labelled accordingly. Within this restricted model the minimum lies at $T = 12$ ns at $6.1\times10^{-5}$.

Within the restricted error model of Figure 4 the minimum lies at a total duration of 12 ns with a ramp duration of 4.50 ns. The range of validity must be stated explicitly at this point, because it separates the two calculations reported here. The half-integer sequence was found to describe the weak-drive minima over the tested range $\Omega/|\alpha| \leq 0.05$; this value is an observed boundary of the present simulations and not a derived constant. The range examined was $\Omega/|\alpha|$ between 0.03 and 0.55; for $\Omega/|\alpha| \leq 0.05$ exactly three minima are found in the interval $\tau = 2$–14 ns and they follow the sequence, while between $\Omega/2\pi = 15$ and 20 MHz, that is $\Omega/|\alpha| \simeq 0.05$–0.06, additional minima appear and a single frequency no longer describes them. At fixed pulse area $\pi$ the weak-drive condition corresponds to $\Omega/2\pi \leq 15.5$ MHz and hence to $T - \tau \geq 32.3$ ns, that is to total durations above approximately 37 ns. The minimum of Figure 4 lies outside this range: there $\Omega/2\pi = 66.7$ MHz and $\Omega/|\alpha| = 0.215$. It is located numerically and is not predicted by the weak-drive sequence. Figure 4 should therefore be read as a breakdown test of the weak-drive spectral-zero picture rather than as an independent analytic prediction. The two values differ measurably: at $T = 12$ ns the computed leakage is $1.01\times10^{-7}$ at $\tau = 4.50$ ns and $5.58\times10^{-5}$ at the half-integer value $\tau = 4.84$ ns, a factor of 550. Below approximately 11 ns the geometric constraint $\tau \leq T/2$ prevents the pulse from accommodating a ramp duration near the first weak-drive half-integer spectral zero, and the leakage rises by orders of magnitude.

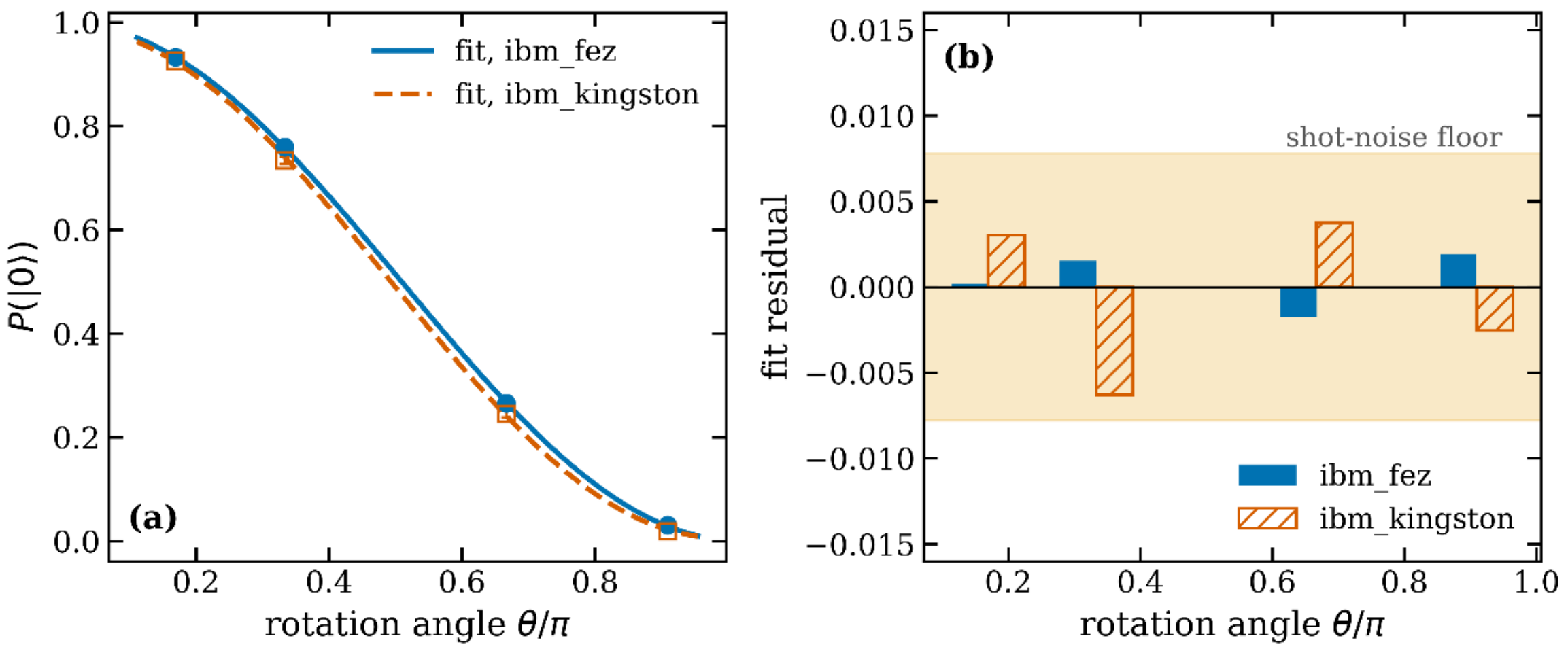

Figure 5. Rotation-angle sweep on ibm_fez and ibm_kingston with 4096 shots per point. (a) Return probability P(|0⟩) and the two-parameter fit of over-rotation and the symmetric readout-offset parameter; error bars are binomial. (b) Residuals of the same fit against the shot-noise band of ±0.0078 at P = 0.5. The residuals lie within the statistical floor, so the four points are fully described by a systematic over-rotation of −1.84 % on ibm_fez and +1.37 % on ibm_kingston with fitted symmetric readout-offset parameters $r = 0.06$ % and 0.69 %, respectively.

A cross-backend comparison was performed using two independent 156-qubit IBM Quantum processors, ibm_fez and ibm_kingston, at four values of the scan parameter with 4096 shots per point.

Fitting the model $P(0) = (1 - 2r)\cos^2[\theta(1+\varepsilon)/2] + r$ to the four points gives a systematic over-rotation $\varepsilon = -1.84$ % with fitted symmetric readout-offset parameter $r = 0.06$ % on ibm_fez, and $\varepsilon = +1.37$ % with $r = 0.69$ % on ibm_kingston. The largest residuals are 0.0019 and 0.0063 respectively, at or below the shot-noise standard deviation of 0.0078 at $P = 0.5$ with 4096 shots. The fitted offset $r$ is a different quantity from the single-point ground-state return error reported above, which was 0.29 % and 0.33 % and includes preparation as well as readout contributions. Since $\eta_{\mathrm{ref}} \propto 1/\Omega_0$ at fixed envelope shape, the measured over-rotations correspond to equivalent fractional shifts in $\eta$ of about 1.4–1.8 %.

At the largest value of the scan parameter sampled in the hardware measurements the return probability is highest. Because the sampled points lie on a monotonic curve, and because the non-adiabaticity parameter and the rotation angle are not independent within the envelope family used for those circuits, the data do not establish a localised resonance feature.

The longitudinal record shows that the measured profiles are sensitive to the calibration state of the device at the time of execution. On ibm_kingston an amplitude audit on 21 April 2026 identified a systematic drive-amplitude shift of approximately 4 % relative to the previous day, consistent with the recalibration recorded in the device logs, and a replication study on five qubits the same day returned values consistent with the new calibration state rather than with the old one.

## 4. Discussion

The principal analytic result is the exact half-integer spectral-zero condition $\tau|\alpha| = n + 1/2$ for a symmetric flat-top envelope with cosine ramps, together with its origin as destructive interference between the contributions of the rising and falling edges. In the weak-drive regime these zeros provide a leading-order prediction for the positions of the endpoint-leakage minima of the multilevel dynamics; they are not an exact cancellation condition for the leakage itself. The envelope-degeneracy relation $\eta_{\mathrm{ref}} \cdot \theta = C_{\mathrm{shape}}$ of Section 2 is the methodological step that makes the result reachable: within a self-similar one-scale family a scan in the non-adiabaticity parameter is indistinguishable from a scan in pulse area, so an independent ramp timescale must be introduced before the ramp duration can be varied at all. One consequence is worth recording separately: for a Gaussian envelope the values $\eta_{ref} \simeq 0.5$ and $\eta_{ref} \simeq 1$, which might be read as thresholds, correspond precisely to a π pulse and a π/2 pulse, since $C_{shape}/\pi = 0.484$.

At the minimum with $\tau|\alpha| \simeq 5/2$, at a total duration of 60 ns and fixed pulse area π, the leakage falls to $5.10\times10^{-9}$ against $2.00\times10^{-6}$ at $\tau = 6.6$ ns, a suppression by a factor of 392.

The half-integer sequence is a weak-drive result: over the range tested it describes the minima for $\Omega/|\alpha| \leq 0.05$. The restricted error-model optimum of Figure 4 lies outside this regime, at $\Omega/|\alpha| = 0.215$, and is located numerically rather than by the analytic condition. At that point the half-integer prediction gives a leakage 550 times larger than the numerical optimum (Section 3), providing a quantitative negative control on extrapolation of the weak-drive sequence into the short-pulse regime.

Near this weak-drive operating point the leakage is substantially less sensitive to amplitude error than the rotation angle is, which suggests that the ramp duration and the drive amplitude can serve as approximately separable local calibration coordinates. No global independence of the two channels is implied.

Because the spectral-zero condition constrains the ramp duration alone, the total pulse duration remains free and can be chosen on other grounds, such as the decoherence budget. To our knowledge the closed-form sequence $\tau|\alpha| = n + 1/2$ for the symmetric flat-top envelope with cosine ramps, determined by the ramp duration alone and independent of the plateau length, has not previously been stated. The general mechanism it belongs to is established: endpoint leakage is governed by the drive spectrum evaluated at the anharmonicity [5], leakage suppression by destructive interference is a known design principle, and the accessible duration range reproduces a previously reported anharmonicity-limited boundary [6]. The envelope-degeneracy relation of Section 2 is a scaling identity and is presented as the methodological step required to isolate the ramp duration, not as a priority claim. A methodological point follows from the verification procedure. At $T = 50$ ns the condition $\tau|\alpha| = n + 1/2$ is numerically indistinguishable from $(T - \tau)|\alpha| = n$, because $50|\alpha| = 15.5$; a sequence inferred from a single total duration would therefore have taken the wrong form. The two were separated only by varying the parameter that the correct condition does not contain.

The present analysis does not improve on established leakage-suppression methods. Frequency-domain envelope shaping achieves a leakage error below $3.0\times10^{-5}$ at a gate duration of 6.25 ns for an anharmonicity of −212 MHz, and an error per gate of $(1.56 \pm 0.07)\times10^{-4}$ at 7.9 ns, using a more sophisticated spectral construction than the flat-top family employed here.

## 5. Experimental accessibility and limitations

Three statements bound what the hardware measurements reported above can and cannot support. First, the predicted operating window is 0.40 ns wide at the ten-fold level, against a control-electronics sample granularity of $dt = 1/4.5$ GHz = 0.222 ns, so the operating point is addressable in timing terms in principle. Second, the measured drive-amplitude calibration biases of −1.84 % and +1.37 % correspond to equivalent fractional shifts in $\eta$ of about 1.4–1.8 %, at which the computed leakage minimum is perturbed by less than 10 %, so the condition is not fragile with respect to the dominant measured systematic. Third, the transfer function of the control chain is not modelled here; it smears sharp ramps and has been identified elsewhere as a dominant error source for gates below 10 ns, so the quoted window is an upper bound on the achievable tolerance. The hardware measurements therefore establish the calibration precision available on these processors and do not constitute a validation of the leakage minima, which were not measured.

## 6. Conclusion

We have derived an exact half-integer sequence of spectral zeros for symmetric flat-top transmon pulses with cosine ramps, $\tau|\alpha| = n + 1/2$, arising from destructive interference between the contributions of the rising and falling edges. In the weak-drive regime these zeros provide a leading-order prediction for the positions of the endpoint-leakage minima. The spectral-zero sequence contains no fitted parameters, and because the cancellation is independent of the plateau length the total pulse duration does not enter. Across six total pulse durations between 50 and 100 ns the computed minima collapse onto the sequence with a root-mean-square deviation of 0.034 in units of $\tau|\alpha|$; for the four durations resolved on the finer grid, inclusion of the first-order population weighting reduces the positional deviation from 0.0234 to 0.0096. The competing condition based on plateau length was excluded at the durations where the two predictions separate. The simulations follow the weak-drive sequence for $\Omega/|\alpha| \leq 0.05$ over the range tested, and at $\Omega/|\alpha| = 0.215$ the prediction fails by a factor of 550 in leakage, which serves as a negative control on the perturbative interpretation. The preceding envelope-degeneracy relation establishes why an independent ramp timescale is required before such a condition can be tested at all. Rotation-angle measurements on two IBM Heron r2 processors give drive-amplitude calibration biases of −1.84 % and +1.37 %, equivalent to fractional shifts in $\eta$ of about 1.4–1.8 %, and a predicted operating window of 0.40 ns that exceeds the 0.222 ns timing granularity of the control hardware; no hardware measurement of the leakage minima is claimed.

These measured biases perturb the computed leakage minimum by less than 10 %, while the hardware record shows day-to-day calibration shifts at the percent level, so practical implementation would require re-verification after each recalibration. The analytic condition is derived in the resonant weak-drive picture, whereas the numerical validation retains the counter-rotating terms; the agreement therefore does not rely on imposing the rotating-wave approximation in the simulation. The half-integer sequence thus provides an exact spectral skeleton for this pulse family, while the displacement of the dynamical minima is governed by the population-weighted pulse geometry.

## Data and code availability

The numerical scripts that generate every computed result and figure reported here, the resulting data in comma-separated form, the hardware execution records with their job identifiers, and the Supplementary Information are openly available in the accompanying GitHub repository: https://github.com/alexandermtishin-art/Half-Integer-Spectral-Zeros-for-Leakage-Suppression-in-Fast-Transmon-Pulses . The repository contains the four-level Duffing model and the leakage scan, the closed-form Fourier amplitude of the envelope with its verification of the half-integer zero condition, the population-weighted first-order amplitude, the envelope shape constants, and the figure-generation scripts, together with a description of the model parameters and of the stated range of validity. The hardware records are provided as a calibration log; they do not contain a measurement of the leakage minima, which were not measured on hardware.

## Competing interests

The author declares no competing interests.

## Declaration of Generative AI in Scientific Writing

The author used AI-based tools (Google Gemini, ChatGPT and Claude) to improve the manuscript's language and to assist with the generation of Python scripts. After using these tools the author reviewed and edited the content as needed and takes full responsibility for the content of the publication.

**References**

1. Chen, Z., Kelly, J., Quintana, *C*., et al. Measuring and suppressing quantum state leakage in a superconducting qubit. Phys. Rev. Lett. 116, 020501 (2016). DOI 10.1103/PhysRevLett.116.020501

2. Gambetta, J. M., Motzoi, *F*., Merkel, S. *T*., and Wilhelm, *F*. K. Analytic control methods for high-fidelity unitary operations in a weakly nonlinear oscillator. Phys. Rev. A 83, 012308 (2011). DOI 10.1103/PhysRevA.83.012308

3. Motzoi, *F*., Gambetta, J. M., Rebentrost, *P*., and Wilhelm, *F*. K. Simple pulses for elimination of leakage in weakly nonlinear qubits. Phys. Rev. Lett. 103, 110501 (2009). DOI 10.1103/PhysRevLett.103.110501

4. Hyyppä, E., Vepsäläinen, A., Papič, M., et al. Reducing leakage of single-qubit gates for superconducting quantum processors using analytical control pulse envelopes. PRX Quantum 5, 030353 (2024). DOI 10.1103/PRXQuantum.5.030353

5. Yang, *H*., Liu, *F*., Wang, W., Fei, Y., and Shan, Z. Separate control of transient leakage exposure and endpoint leakage in fast transmon gates. arXiv:2607.05779 (2026). DOI 10.48550/arXiv.2607.05779

6. Türkpençe, D., and Çakmak, S. Anharmonicity-limited gate-time boundaries in transmon single-qubit quantum gates. J. Opt. Soc. Am. B 43, 245–254 (2026). DOI 10.1364/JOSAB.581099

7. Schutjens, R., Abu Dagga, F., Egger, D. J., and Wilhelm, F. K. Single-qubit gates in frequency-crowded transmon systems. Phys. Rev. A 88, 052330 (2013). DOI 10.1103/PhysRevA.88.052330

8. Malarchick, *R*. When does numerical pulse optimization actually help? Error budgets, robustness tradeoffs, and calibration guidance for transmon single-qubit gates. arXiv:2511.12799 (2026). DOI 10.48550/arXiv.2511.12799

9. Lazăr, S., Ficheux, Q., Herrmann, J., Remm, A., Lacroix, N., Hellings, C., Swiadek, F., Colao Zanuz, D., Norris, G. J., Bahrami Panah, M., Flasby, A., Kerschbaum, M., Besse, J.-C., Eichler, C., and Wallraff, A. Calibration of drive non-linearity for arbitrary-angle single-qubit gates using error amplification. Phys. Rev. Applied 20, 024036 (2023). DOI

10.1103/PhysRevApplied.20.024036

10. Werninghaus, M., Egger, *D*. J., Roy, *F*., Machnes, S., Wilhelm, *F*. K., and Filipp, S. Leakage reduction in fast superconducting qubit gates via optimal control. npj Quantum Inf. 7, 14 (2021). DOI 10.1038/s41534-020-00346-2

11. Tishin, A. M. An effective scaling framework for non-adiabatic mode dynamics. Chaos, Solitons and Fractals 211, 118839 (2026). DOI 10.1016/j.chaos.2026.118839